\documentclass[aps,11pt,superscriptaddress,nofootinbib]{revtex4-1}
\usepackage{amsfonts}
\usepackage{amsmath}
\usepackage{amssymb}
\usepackage{graphicx}
\def\Title#1{\begin{center} {\Large {\bf #1} } \end{center}}

\begin{document}

\Title{Anisotropic stellar structure equations for magnetized  stars}

\bigskip\bigskip


\author{D. Manreza Paret}
\affiliation{Departamento de F\'{\i}sica General, Facultad de F\'{\i}sica, Universidad de la Habana, La Habana, 10400, Cuba}
\affiliation{Instituto de Astronomia, Geofisica e Ciencias Atmosfericas USP, Rua do Matao 1226, 05508-900 Sao Paulo SP, Brazil}
\author{J. E. Horvath}
\affiliation{Instituto de Astronomia, Geofisica e Ciencias Atmosfericas USP, Rua do Matao 1226, 05508-900 Sao Paulo SP, Brazil}
\author{A. P\'{e}rez Mart\'{\i}nez}
\affiliation{Instituto de Cibern\'{e}tica, Matem\'{a}tica y F\'{\i}sica (ICIMAF),\\
 Calle E esq a 15 No 209, Vedado, La Habana, 10400, Cuba}

\begin{abstract}
The fact that a magnetic field in a fermion system breaks the spherical symmetry suggest that the intrinsic geometry of this system is axisymmetric rather than spherical. In this work we analyze the impact of anisotropic pressures, due to the presence of a magnetic field, in the structure equations of a magnetized quark star. We assume a cylindrical metric and an anisotropic energy momentum tensor for the source. We found that there is a maximum magnetic field that the star can sustain, closely related to the violation of the virial relations.
\end{abstract}
\maketitle

\section{Introduction}
The presence of huge magnetic fields in compact objects (CO) is an established fact. Typical measured surface magnetic fields are about $10^{12}$ G, although in the case of the {\it magnetar} subclass they can be as high as $10^{15}$ G \cite{Woods2006csxs}. However, there are no observations capable of measuring the magnetic fields in the inner regions of the stars, and theoretical arguments must be used to estimate maximum values of the field with an eye on possible modifications to the stellar tructure. Based on scalar virial theorem \cite{Shapiro1991ApJ} it can be estimated that the maximum magnetic field that a CO can sustain as $(4\pi R^3/3)(B_{\text{max}}/8\pi)\sim GM^2/R\,\,\Rightarrow\,\,B_{\text{max}}\sim2\times10^8(M/M_\odot)(R/R\odot)^{-2}$ G, for a neutron star whit a mass $M = 1.4M_\odot$ and a radius $R \sim 10^{6} cm = 10^{-4} R_\odot$, where $M_\odot$ and $R_\odot$ are the mass and the radius of the sun respectively, we find that $B_{\text{max}}\sim10^{18}$ G. For a self--bound star a higher maximum value inside the core  $B_{\text{max}}\sim10^{20}$ G has been suggested in Ref.\cite{Ferrer2010PhRvC}.
These numbers suggest the idea that a realistic model of a CO must consider that matter is magnetized, and a few attempts have been done in order to construct models that describe the microphysics magnetized fermion systems \cite{Felipe2005ChJAA,Chakrabarty1996PhRvD}. 

In the simplest case in which the source of the magnetic field is not addressed, i.e. fermions in an external field case, it is quite clear that in fermion systems the former breaks the spherical symmetry and produces an anisotropy in the pressure. Depending on its actual numerical value, this anisotropy could induce a deformation of the CO, and eventually leads to an anisotropic collapse of the object \cite{Aurora2003EPJC} if ultra strong magnetic fields are indeed present. The limits of the absolute maximum field and the issue of spherical symmetry of the COs are the subjects of the present paper.

Anisotropic systems have been studied in the context of stars, but the standard approach has been to take spherical symmetry for granted \cite{Dev2000astro.ph,Harko2003RSPSA}. In the presence of pressure anisotropies due to a magnetic field other symmetry choices (i.e. cylindrical symmetry) could give a more complete description of the physics, this feature has been pointed out by several authors \cite{Paulucci2011PhRvD, Dexheimer2012EPJA}. Recently some works dealing with the problem of axisymmetric metric \cite{Herrera2013PhRvD,Quevedo2012arXiv1201} have been presented, although they remained within a theoretical perspective without application to actual systems.

Given these arguments, the introduction of a {\it cylindrical} symmetric metric in the Einstein equation, together with the construction of an anisotropic energy momentum tensor for the magnetized matter seams a more ``natural'' choice. This anisotropic hydrostatic equilibrium equation could shed some light about how the magnetic field affects the sphericity of the CO, and yield upper limits for the values of the magnetic field that this objects can sustain.

In a first approximation to investigate this problem we will use a general cylindrical symmetric metric, whit coordinates ($t,r,\phi,z$), following the procedures of \cite{Trendafilova2011EJPh} to solve Einstein equations for an axisymmetric model of a CO to take into account the anisotropy induced by the external magnetic field. In the presence of a constant, external magnetic field there are two main directions in space, parallel and perpendicular to the magnetic field. One of the main approximations that we shall make is that all the functions and variables of our model will depend on the radial $(r)$ variable only, so that we can simply describe the perpendicular (equatorial) direction of the CO with respect to the magnetic field. We also consider that the magnetic field is constant and in the $z$ direction. This model can give information of the effects of the magnetic field in terms of the shape (oblateness) of the CO and yield upper limits for the values of the magnetic field that this objects can sustain. More realistic models will be studied in the future.

For the microscopic description of magnetized matter we use the results obtained in Ref. \cite{Aurora2010IJMPD,Felipe2009JPhG} for magnetized strange quark matter (MSQM)  which can correspond to the composition of a strange quark star in its self-bound version, or to the central nucleus of a hybrid neutron star if it is favored at high pressure only. Both possibilities can be achieved by selecting different sets of the available parameters, as discussed in Ref. \cite{Dexheimer2012EPJA,Paulucci2011PhRvD}.

In section \ref{sec1} we review the main thermodynamical properties of a magnetized quark gas used to describe the matter inside the star. In section \ref{sec2} we obtain the mass-radius relation for MSQM in a standard spherical symmetry first and show the problems related to the existence of two pressures; next Einstein equations are solved in cylindrical symmetry to obtain the structure equations that describe the equilibrium of the stars taking into account both pressures. Finally in section \ref{sec4} we present the conclusions of this study.

\section{Magnetized fermion system}\label{sec1}

The thermodynamical potential for a magnetized fermion gas in the framework of the MIT Bag
Model is given by
\begin{equation}\label{OVolumen}
\Omega_f(\mu_f,B,T)= -\frac{d_fe_fB}{\beta}\left[ \sum_{l=0}^{\infty}\sum_{p_4}\int\limits_{-\infty}^{\infty}\frac{dp_3}{(2\pi)^2} \ln \det G^{-1}_{f}(\overline{p}^*)\right],
\end{equation}
with ${\overline{p}}^{*}=(ip^{4}-\mu_f,0, \sqrt{2e_fBl},p^{3})$ for $l=0,1,2,...$;  $\beta$ is the inverse absolute temperature, $\mu_f$ is the fermionic chemical potential and  $G^{-1}_{f}=det[\overline{p}^{*}\cdot\gamma-m_f]$.

We label with $f$ the electrons and $\,u, \,d, \,s$ the quark flavors.
After performing the Matsubara sum we obtain
\begin{equation}
\Omega_f(B,\mu_f,T)=-\frac{d_fe_fB}{2\pi^2}\left[\sum_{l=0}^{\infty}\alpha_l\int\limits_{0}^{\infty}dp_3 \left (\mathcal{E}_f +\frac{1}{\beta}ln(1+e^{-(\mathcal{E}_f-\mu_f)})(1+e^{-(\mathcal{E}_f+\mu_f)})\right )\right]
\end{equation}

Taking the zero temperature limit (compact objects are considered highly degenerate ($\mu \gg T$), therefore the thermal effects can be neglected) we can write the thermodynamic potential as a sum of the vacuum and statistical contributions
\begin{equation}\label{Omega1}
  \Omega_f= \Omega_f(B,0,0) + \Omega_f(B,\mu,0).
\end{equation}
with
\begin{equation}\label{Omegav}
 \Omega_f(B,0,0)= -\frac{e_fB}{4\pi^2}\sum_{l=0}^{\infty}\int dp_3|\mathcal{E}_{f}|,
\end{equation}

\noindent where $\mathcal{E}_f=\sqrt{p_3^2+m_f^2 +2|e_f B|l}$. The $\Omega_f(B,0,0)$ is the vacuum contribution and  the renormalized form was found in \cite{Berestetskii1980MINTF}. In what follows we neglect it, since
we are interested here in a region of fields $eB \leq \mu^2$  where the statistical contribution $\Omega_f(B,\mu_f,0)$ to the physical quantities
is more important than the vacuum one.

The statistical contribution $\Omega_f(B,\mu_f,0)$ has the form
\begin{eqnarray}\label{OM}
   \Omega_f(B,\mu_f,0) =-\frac{d_fe_fB}{2\pi^2}\sum_{l=0}^{l_{max}}\alpha_l\int_0^{\sqrt{\mu_{f}^2-\varepsilon_f^2}} dp_3\mu_f\sqrt{p_{3,f}^2+\varepsilon_f^2}\\
  = -\frac{d_fe_fB}{4\pi ^2}\left [\sum_{l=0}^{l_{max}}\alpha_{l}\left( \mu_f\,{p}_{F}-{\varepsilon}_{f}^2\ln\frac{ {\mu_f} + {p}_{F}}{{\varepsilon}_{f}}\right)\right ],
\end{eqnarray}
where $l_{max}= [\frac{\mu_f^2-m_f^2}{2eB}]$, $I[z]$ denotes the integer part of $z$, $\alpha_{l}=2-\delta_{l0}$ is the spin degeneracy of the $l$-
Landau level and  $d_e=1$ and $d_{u,d,s}=3$ are degeneracy factors.
The Fermi momenta  is  ${p}_F=\sqrt{{\mu_f}^2-{\varepsilon}_{f}^2}$ and  the rest energy is given as
\begin{equation}
\varepsilon_{f}=\sqrt{m_f^2 +2|e_f B|l},
\end{equation}

For degenerate magnetized strange quark matter, the number density and magnetization are given by the expressions
\begin{eqnarray}
N_f&=&-(\partial\Omega_f/\partial \mu_f)=\frac{d_fm^2}{2\pi^2}\frac{B}{B^c_{f}}\sum_{l=0}^{l_{max}}\alpha_{l}{p}_F,\label{Density1}\\
\mathcal{M}_f&=&-(\partial\Omega_f/\partial B)=\frac{d_fe_f m_f}{4\pi ^2}\left (\sum_{l=0}^{l_{max}}\alpha_{l}\left[{\mu_f}{p}_F-\left[ {\varepsilon}_{f}^2 + 2{\varepsilon}_{f} {C}_f \right]\ln\frac{{\mu_f}+ {p}_F}{{\varepsilon}_{f}}\right]\right),\label{Magnetizacion}
\end{eqnarray}
where $B^{c}_f=m_{f}^2/|e_f|$ is the critical
magnetic field  and ${C}_f =\frac{\frac{B}{B_f^c}l}{\sqrt{2l\frac{B}{B_f^{c}}+m_f^2}}$.

Energy density and pressures are
\begin{subequations}
\begin{eqnarray}
   \epsilon&=& \Omega_f + \mu_f N_f,\label{EnerPresure1a}\\
  \mathcal{P}_{\parallel}  &=&-\sum_f\Omega_f,\label{EnerPresure1b}\\
   \mathcal{P}_\bot &=&\sum_f(-\Omega_f-B\mathcal{M}_f),\label{EnerPresure1c}
\end{eqnarray}
\end{subequations}
In order to study the matter inside the star we use the MIT Bag model, thus the equation of state is obtained from (\ref{EnerPresure1a}), (\ref{EnerPresure1b}) and (\ref{EnerPresure1c}) adding the bag (vacuum energy) parameter and the classical magnetic energy.,
\begin{eqnarray}
  E &=& \varepsilon+\frac{B^2}{8\pi}+B_{bag}, \label{EoS1}\\
  P_{\parallel} &=& \mathcal{P}_{\parallel}-\frac{B^2}{8\pi}-B_{bag}, \label{EoS2}\\
  P_{\perp} &=& \mathcal{P}_{\perp}+\frac{B^2}{8\pi}-B_{bag}.\label{EoS3}
\end{eqnarray}

The stelar chemical equilibrium conditions are obtained solving the system of equations
\begin{subequations}\label{consNb}
\begin{eqnarray}
\mu_u+\mu_e-\mu_d=0, \,\,\, \,\,\, \mu_d-\mu_s&=&0,\ \ \  \beta\,\text{equilibrium},\\
2N_u-N_d-N_s-3N_e&=&0,\ \ \   \text{charge neutrality}, \\
N_u+N_d+N_s-3 n_{B}&=&0\ \ \   \text{baryon number conservation}.
\end{eqnarray}
\end{subequations}

Once the system (\ref{consNb}) is solved, we can find the thermodynamical properties of the MSQM in stellar equilibrium conditions and study how the magnetic field modified them.
\begin{figure}[!ht]
\centering
      \includegraphics[height=7.0cm,width=10cm]{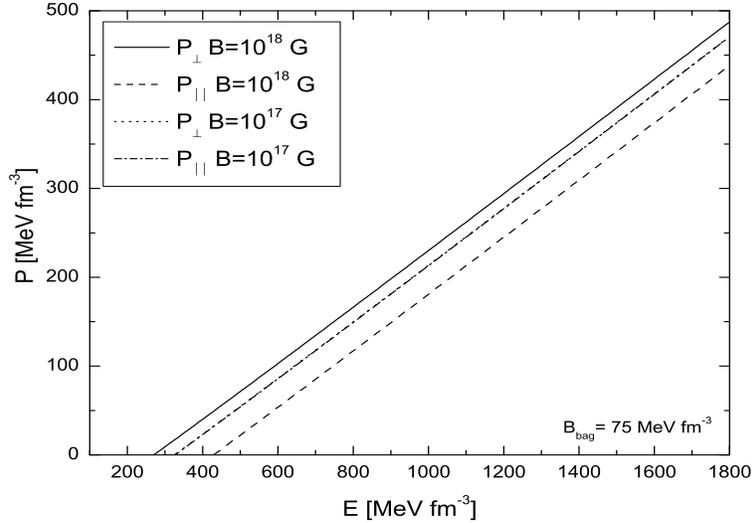}
      \caption{Equations of state for magnetized strange quark matter.}
      \label{fig1}
\end{figure}
\begin{figure}[!ht]
\centering
      \includegraphics[height=5.0cm,width=7cm]{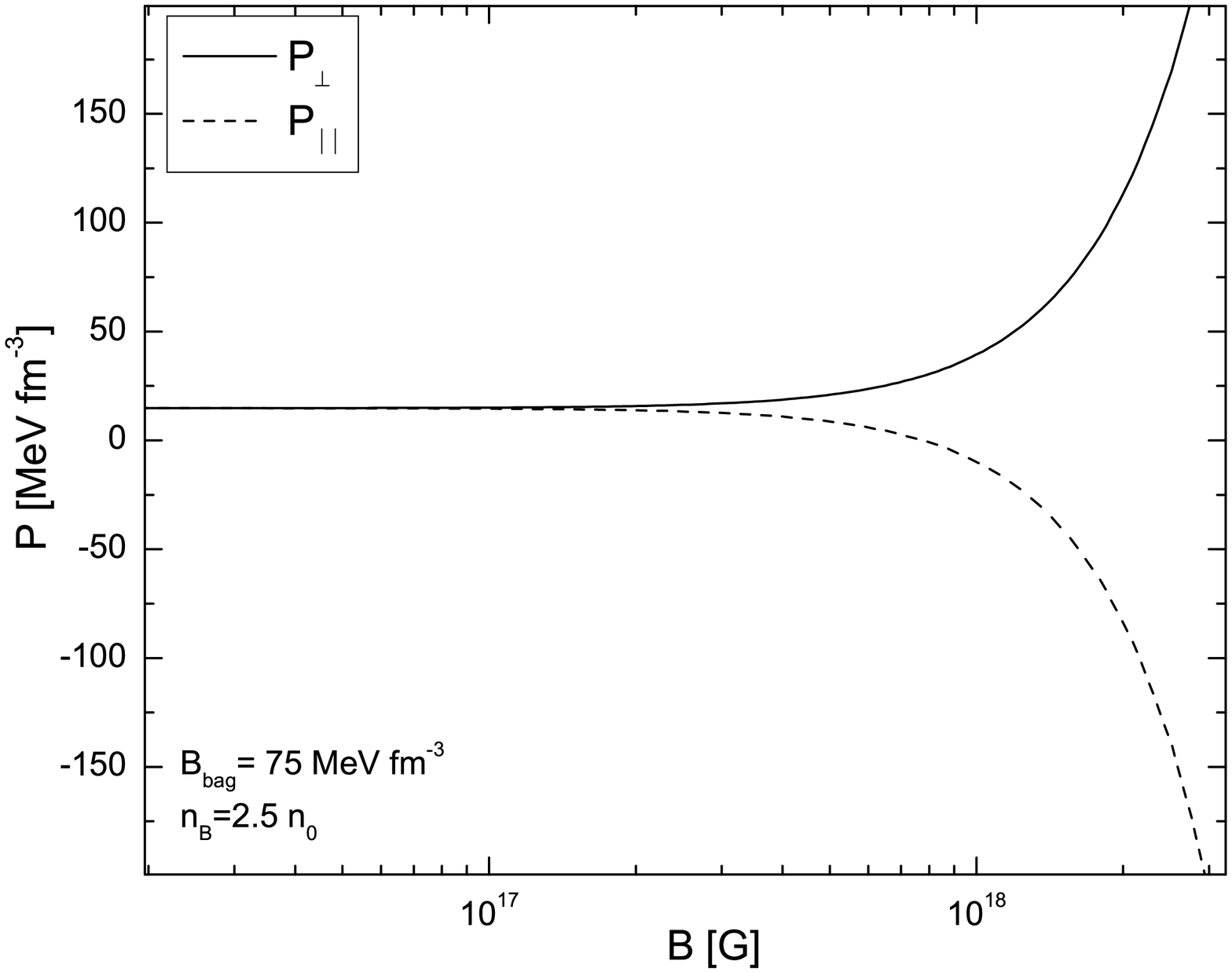}
      \includegraphics[height=5.0cm,width=7cm]{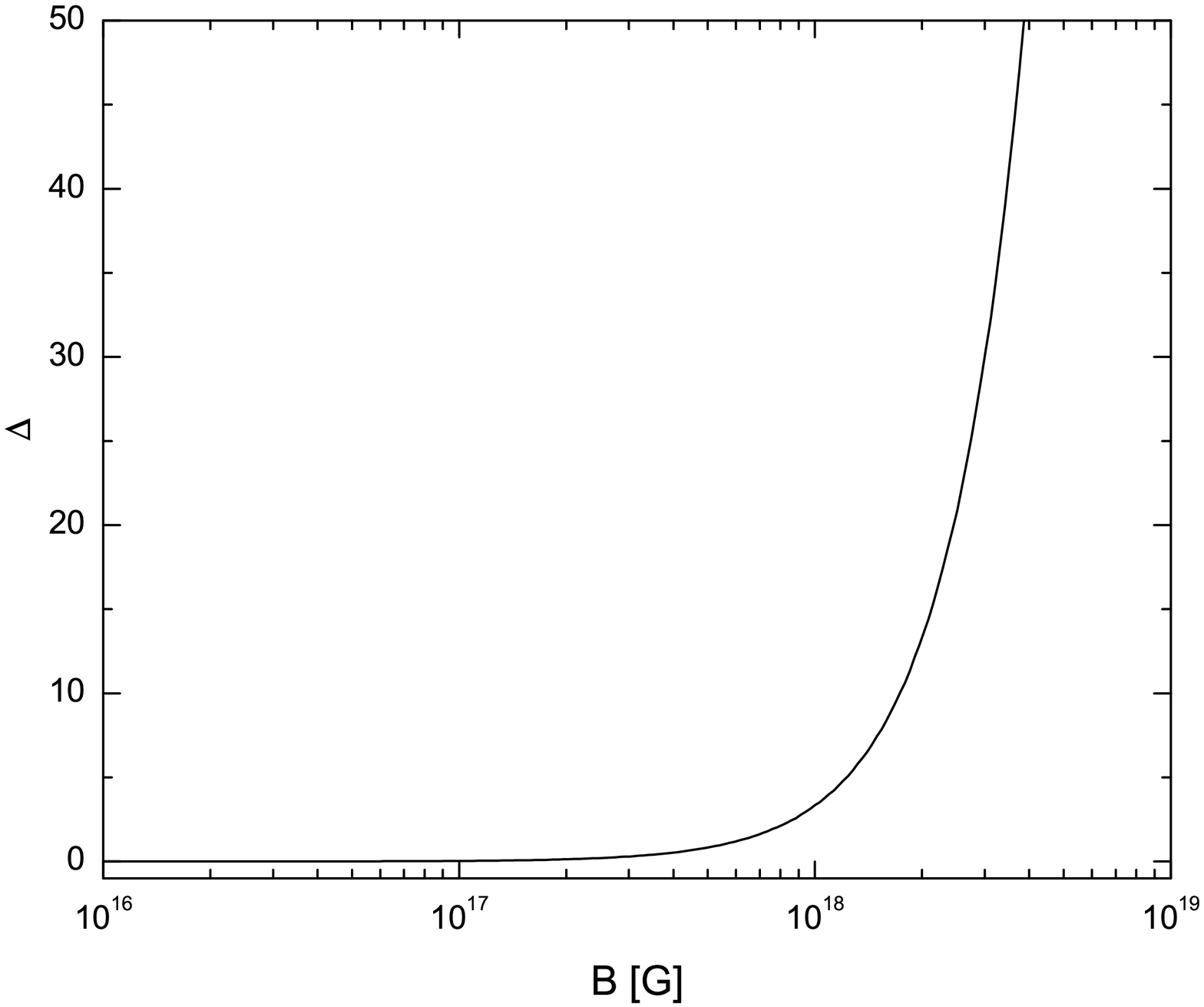}
      \caption{Dependence of the pressures and the splitting coefficient with respect to the magnetic field.}
      \label{fig2}
\end{figure}

In Fig.\,\ref{fig1} we show the EOS of the magnetized gas, stressing the fact that when we increase the magnetic field the anisotropy becomes relevant. An even more illustrative graphic is the dependence of pressures on the magnetic field which we show in Fig.\,\ref{fig2}. It is noted that when the magnetic field increases, the splitting of the pressures becomes greater, as expected. There is a regime where the pressures are nearly equal (isotropic regime), but for fields around $B\sim10^{18}$ G the pressure anisotropy becomes very large. A quantitative parameter to measure the importance of the pressure anisotropy (the {\it splitting coefficient}) can be defined as
\begin{equation}\label{split}
    \Delta=\frac{|P_{\perp}-P_{\parallel}|}{P(B\rightarrow0)}.
\end{equation}
In the right panel of Fig.\,\ref{fig2} we can see the dependence of this coefficient with the magnetic field. A criterion to discriminate between isotropy and anisotropy regimes is that $\Delta\simeq\mathcal{O}(1)$ \cite{Ferrer2010PhRvC}. In our case $\Delta=1$ for a magnetic field $B=5\times10^{17}$ G, while for $B=10^{18}$ G $\Delta\simeq3.3$. In our numerical computations we will first use magnetic field values well within the isotropic region $B=10^{17}$ G, and after that in the anisotropic region $B=10^{18}$ G to compare their effects on the star structure.

\section{TOV equations for MSQM}\label{sec2}

In order to set up the problem posed by the magnetized matter EoS in the study of the structure of CO, we will analyze the usual, spherical case, solving the resulting TOV equations \cite{Misner1973grav}. To find the static structure of a relativistic spherical star the system is
\begin{equation}
G^{\mu}_{\,\,\,\,\nu}\equiv R^{\mu}_{\,\,\,\,\nu}-\frac{1}{2}R
g^{\mu}_{\,\,\,\,\nu}=8\pi G \mathcal{T}^{\mu}_{\,\,\,\,\nu},
\label{EE1}
\end{equation}
$(\mu,\nu=0,1,2,3)$, using the Schwarzschild metric
\begin{equation}\label{schw1}
  ds^2=-e^{2\nu} dt^2+e^{2\lambda} dr^2+r^2d\Omega^2,\,\,\,\, d\Omega^2=d\theta^2+\sin^2\phi d\phi^2,
\end{equation}
and the energy momentum tensor
\begin{equation}\label{TE-M}
\mathcal{T}^{\mu}_{\,\,\,\,\nu}=(E+P)u^{\mu}u_{\nu}+Pg^{\mu}_{\,\,\nu},
\end{equation}

we obtain the TOV equations
\begin{eqnarray}
  \frac{dM}{dr} &=& 4\pi G\,E, \label{TOV1} \\
  \frac{dP}{dr} &=& -G\frac{(E+P)(M+4\pi Pr^3)}{r^2-2rM}, \label{TOV2}
\end{eqnarray}
with boundary conditions $P(R)=0$, $M(0)=0$ and the EoS $E\rightarrow f(P)$.
\begin{figure}[!ht]
\centering
      \includegraphics[height=7.0cm,width=10cm]{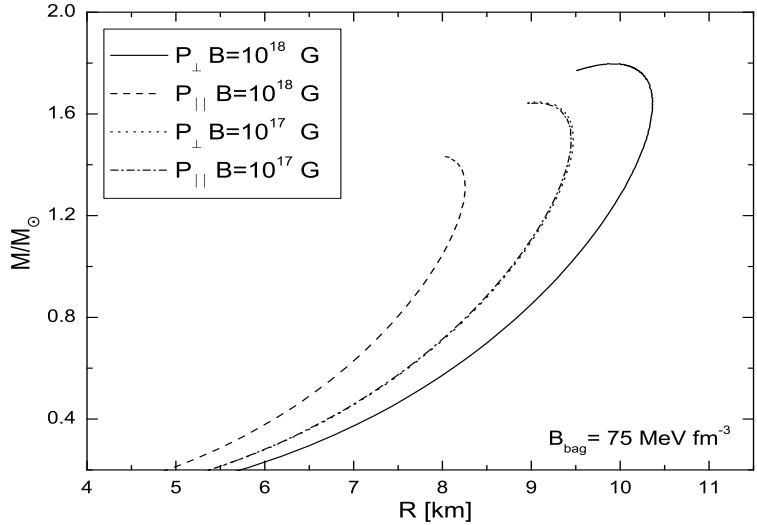}
      \caption{Mass-Radius diagram comparing the effects of taking parallel or perpendicular pressure.}
      \label{fig3}
\end{figure}

When we look at the equation (\ref{TOV2}) the problem of which pressure must be used in the case of a magnetized EoS (like the one obtained in the preceding section) arises. One option is to work within the isotropic regime $(\Delta<1)$ where $P_{\perp}=P_{\parallel}$. However, if we want to explore the anisotropic regime and address the maximum field issue, the problem of is unavoidable.

In Fig.\,\ref{fig3} it is shown the mass--radius diagram for MSQM for two values of the magnetic field, comparing the effects of both pressures in equation (\ref{TOV2}). For $B=10^{17}$ G the differences are not visible as expected because the pressures are nearly equal and we are in the isotropic regime $(\Delta=0.03)$. For $B=10^{18}$ G the differences between using the perpendicular or the parallel pressure are quite large. In this case one most establish a criterion to employ one of them or improve the structure equations to take into account the anisotropies.

\section{Anisotropic structure equations}\label{sec3}

In order to improve the structure equations in presence of anisotropic pressures we propose that a more ``natural'' geometry of a magnetized fermion system is an axisymmetric geometry thus, to obtain the structure equations we start with the cylindrically symmetric metric
\begin{equation}\label{cyl1}
  ds^2=-e^{2\Phi} dt^2+e^{2\Lambda} dr^2+r^2d\phi^2+e^{2\Psi} dz^2
\end{equation}
where $\Phi$, $\Lambda$, $\Omega$, and $\Psi$ are functions of $r$ only.

For this metric, the nonzero Einstein tensor components are
\begin{eqnarray}
G^t_{t}&=&e^{-2\Lambda}(\Psi'' +\Psi'^2 -\Psi'\Lambda' -\frac{1}{r}\Lambda' +\frac{1}{r}\Psi') \nonumber\\
G^r_{r}&=&e^{-2\Lambda}(\Psi' \Phi' +\frac{1}{r}\Phi' +\frac{1}{r}\Psi') \nonumber\\
G^\phi_{\phi}&=&e^{-2\Lambda}(\Phi''+\Phi'^2-\Phi'\Lambda'+\Psi''+\Psi'^2-\Psi'\Lambda'+\Psi'\Phi')\nonumber \\
G^z_{z}&=&e^{-2\Lambda}(\Phi''+\Phi'^2-\Phi'\Lambda'-\frac{1}{r}\Lambda' +\frac{1}{r}\Phi') \nonumber
\end{eqnarray}

With the energy momentum tensor for magnetized mater given by \cite{Felipe2005ChJAA}
\begin{equation} \mathcal{T}^{\mu}_{\,\,\,\,\nu} = \left(
\begin{array}{llll}
E & 0&0&0 \\
0& P_{\perp}&0&0\\
0&0&P_{\perp}&0\\
0&0&0&P_{\parallel}
\end{array}\right), \label{presiones}
\end{equation}
where $E$, $P_{\parallel}$ and $P_{\perp}$ are given by the EoS (\ref{EoS1}), (\ref{EoS2}) and (\ref{EoS3}) respectively.

From the Einstein field equations in natural units we then obtain the following four differential equations:
\begin{eqnarray}
4\pi E&=&-e^{-2\Lambda}(\Psi'' +\Psi'^2 -\Psi'\Lambda' -\frac{1}{r}\Lambda' +\frac{1}{r}\Psi') \nonumber\\
4\pi P_{\perp}&=&e^{-2\Lambda}(\Psi' \Phi' +\frac{1}{r}\Phi' +\frac{1}{r}\Psi') \nonumber\\
4\pi P_{\perp}&=&e^{-2\Lambda}(\Phi''+\Phi'^2-\Phi'\Lambda'+\Psi''+\Psi'^2-\Psi'\Lambda'+\Psi'\Phi') \nonumber\\
4\pi P_{\parallel}&=&e^{-2\Lambda}(\Phi''+\Phi'^2-\Phi'\Lambda'-\frac{1}{r}\Lambda' +\frac{1}{r}\Phi') \nonumber
\end{eqnarray}

Performing some algebra with the previous system of equations, and using the energy momentum conservation $(\mathcal{T}^{\mu}_{\,\,\,\,\nu;\mu})$ we finally obtain
\begin{subequations}\label{Diff2}
\begin{eqnarray}
P_{\perp}'&=&-\Phi'(E+P_{\perp})-\Psi'(P_{\perp}-P_{\parallel})\\
4\pi e^{2\Lambda}(E+P_{\parallel}+2P_{\perp})&=&\Phi''+\Phi'(\Psi'+\Phi'-\Lambda')+\frac{\Phi'}{r} \\
4\pi e^{2\Lambda}(E+P_{\parallel}-2P_{\perp})&=&-\Psi''-\Psi'(\Psi'+\Phi'-\Lambda')-\frac{\Psi'}{r} \\
4\pi e^{2\Lambda}(P_{\parallel}-E)&=&\frac{1}{r}(\Psi'+\Phi'-\Lambda')
\end{eqnarray}
\end{subequations}

This form, together with the EoS $E\rightarrow f(P_{\perp}),\,\, P_{\parallel}\rightarrow f(E)$ is a system of differential equations in the variables
\begin{equation}
  P_{\perp},\,\,\,P_{\parallel},\,\,\,E,\,\,\, \Phi,\,\,\, \Lambda,\,\,\, \Psi,
\end{equation}

Since the differential equations involve factors of $1/r$ , we will start with a power series expansions of $P_{\perp}$, $\Phi$, $\Psi$, and $\Lambda$ around $r=0$ to find initial conditions suitable for numerical calculations
\begin{eqnarray}
P_{\perp}&=&P_{\perp0}+P_{\perp1} r, \label{t1}\\
\Lambda&=&\Lambda_0+\Lambda_1 r, \label{t2}\\
\Phi&=&\Phi_0+\Phi_1 r + \Phi_2 r^2, \label{t3}\\
\Psi&=&\Psi_0+\Psi_1 r + \Psi_2 r^2 \label{t4}.
\end{eqnarray}

We take also $\Psi=\Phi=\Lambda=0$ at $r=0$ so that the corresponding metric coefficients are equal to $1$ at that point and $\Psi'=\Phi'=0$ to select smooth solutions on the $z$-axis.

By substitution of these conditions in the system of differential equations we find
\begin{subequations}\label{IC1}
\begin{eqnarray}
  P_{\perp}(0) &=&P_{\perp0}  \\
  \Lambda(0) &=& 0 \\
  \Phi(0) &=& \frac{1}{2}(P_{\parallel0}+2P_{\perp0}+E_0)(r_0^2-2r_0) \\
  \Psi(0) &=& \frac{1}{2}(-P_{\parallel0}+2P_{\perp0}-E_0)(r_0^2-2r_0) \\
  \Phi'(0) &=& 0 \\
  \Psi'(0) &=& 0
\end{eqnarray}
\end{subequations}
And we also impose
\begin{equation*}
   P_{\perp}(R_{\perp})=0
\end{equation*}
which determines the radius of the star, in the equatorial (perpendicular) direction.

\begin{figure}[!ht]
\centering
      \includegraphics[height=5.0cm,width=7cm]{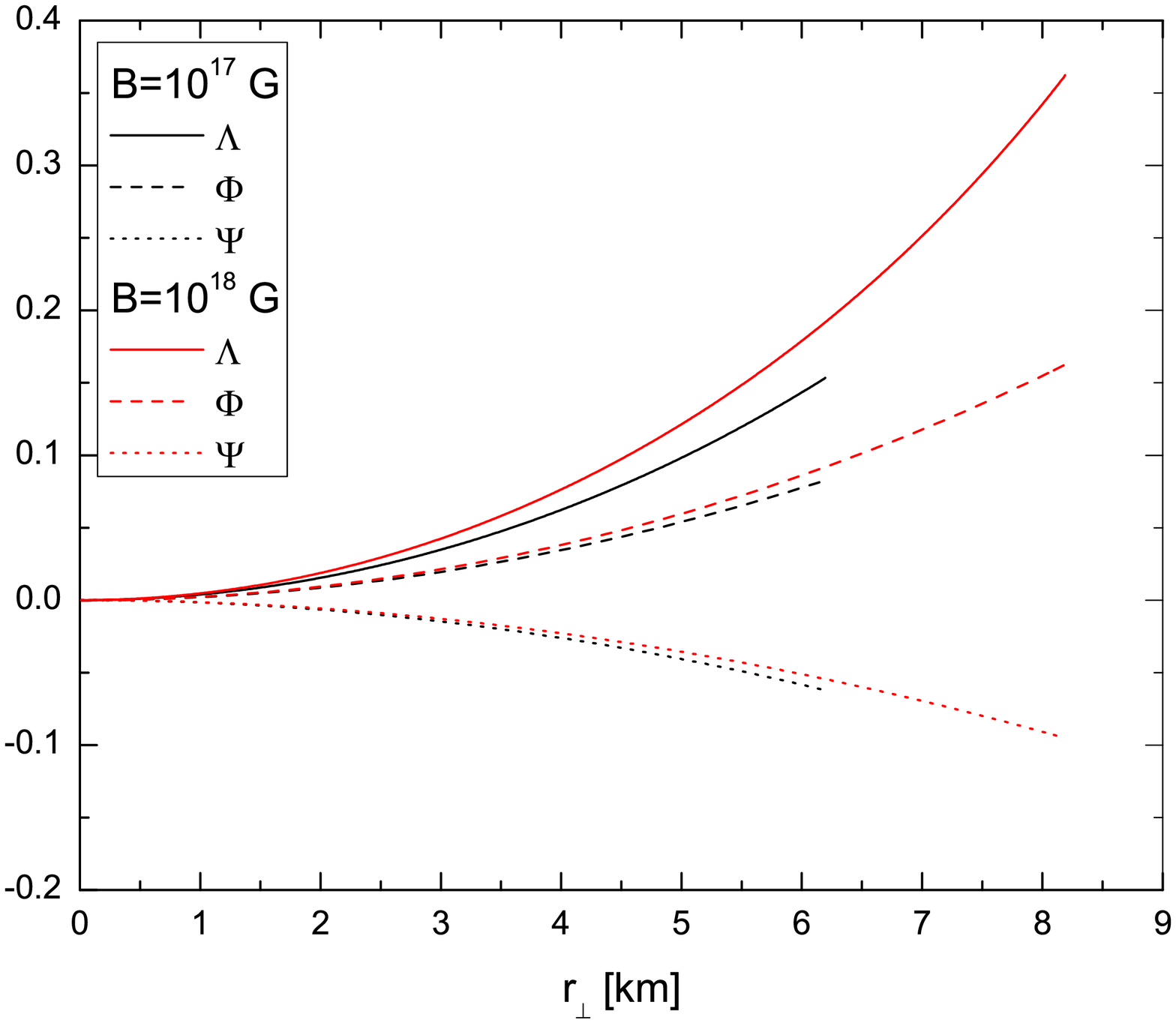}
      \includegraphics[height=5.0cm,width=7cm]{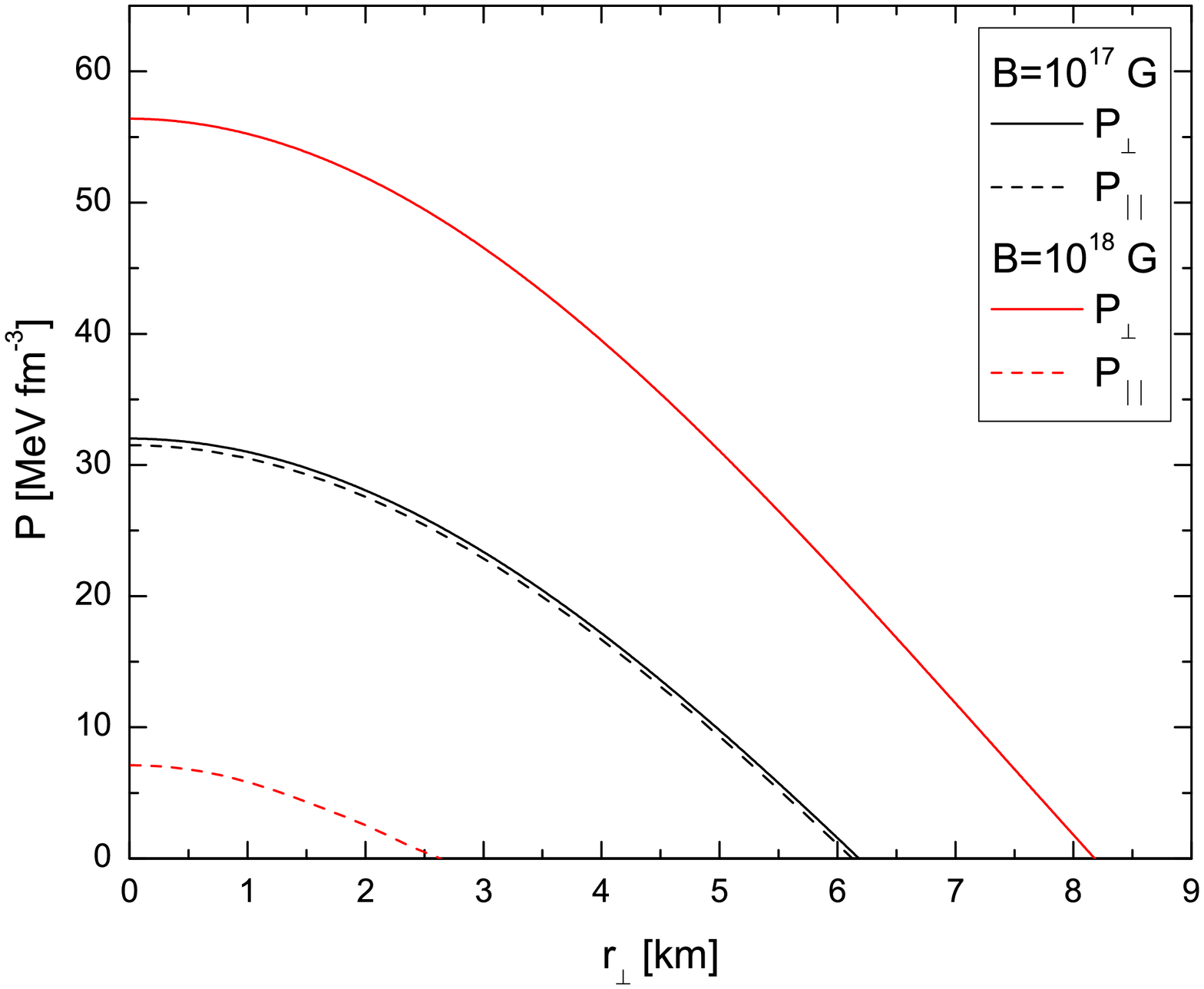}
      \caption{Metric coefficients and pressures inside the star for two values of the magnetic field.}
      \label{fig4}
\end{figure}

The solutions of the system of equations (\ref{Diff2}) with initial conditions (\ref{IC1}) are shown in Fig.\,\ref{fig4}. In the left panel of Fig.\,\ref{fig4} the behavior of the metric coefficients is shown for two values of the magnetic field, as we can note, the increase of the magnetic field produces an increase of the star radius in the perpendicular direction. In the right panel of Fig.\,\ref{fig4} the pressures inside the star are depicted for a selected central density. All this quantities present a regular physical behavior.

As we have pointed out, by hypothesis all our variables depend just on the perpendicular radial direction. Therefore, we can not simply compute total quantities such as the mass radius diagram like Fig.\,\ref{fig3} in the case of spherical symmetry. Instead we will compute the Tolman \cite{Tolman1934rtc} generalization for the the mass per unit length of a source
\begin{equation}\label{tolman1}
  M_T=\int\sqrt{-g}(T^0_0-T^1_1-T^2_2-T^3_3)dV
\end{equation}
for the cylindric metric (\ref{cyl1}) we have
\begin{eqnarray}\label{tlman2}
   M_T&=&\int re^{\Phi+\Psi+\Lambda}(E-2P_{\perp}-P_{\parallel})dV \\
   &=&\int_0^{2\pi}\int_{-R_{\parallel}}^{R_{\parallel}}\int_0^{R_{\perp}} re^{\Phi+\Psi+\Lambda}(E-2P_{\perp}-P_{\parallel})d\phi\,dz\,dr \\
   &=& 4\pi R_{\parallel}\int_0^{R_{\perp}} re^{\Phi+\Psi+\Lambda}(E-2P_{\perp}-P_{\parallel})dr
\end{eqnarray}
Therefore, we can not compute the mass of the star but rather the mass per unit length $(M_T/R_{\parallel})$
\begin{equation}\label{tlman3}
   \frac{M_T}{R_{\parallel}}=4\pi\int_0^{R_{\perp}} re^{\Phi+\Psi+\Lambda}(\epsilon-2P_{\perp}-P_{\parallel})dr
\end{equation}

\begin{figure}[!ht]
\centering
      \includegraphics[height=7.0cm,width=10cm]{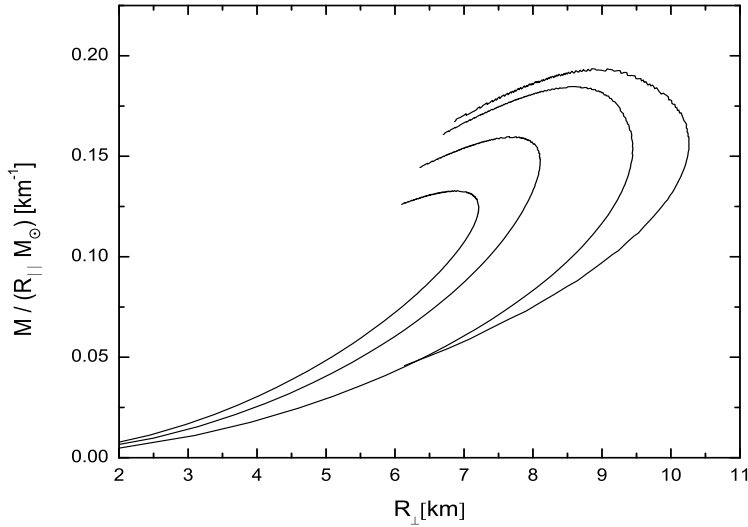}
      \caption{Mass per unit parallel length $(M/R_{\parallel})$ in solar mases, versus perpendicular radius. As the magnetic field increases the perpendicular radius increases up to a critical field. The curves are organized in order of increasing values of the magnetic fields: $B=10^{17}$ G, $B=10^{18}$ G, $B=1.5\times10^{18}$ G and $B=1.7\times10^{18}$ G.}
      \label{fig5}
\end{figure}

In Fig.\,\ref{fig5} the mass per unit length versus perpendicular radius is shown. When the magnetic field increases, the perpendicular radius and the mass per unit length of the star also increase. It is found that there is a maximum field ($B\simeq1.8\times10^{18}$ G) beyond which the metric coefficients exhibit a divergent behavior, this value of the magnetic field almost coincides with the threshold for which the pressure difference has become important, and results $B=1.8\times10^{18}$ G, (for which $\Delta=10.5$). Therefore, we interpret that no stable solutions of the system are possible beyond this point and this signals the end of the theoretical stellar sequences within the adopted assumptions.
Even though within our model we can not compute the mass radius relation the information given in Fig.\,\ref{fig5} its important to constrain the maximum magnetic field allowed for magnetized CO.

\section{Conclusions}\label{sec4}

We have pointed out and worked on the problem of anisotropic pressures in the description of the structure of a CO in this paper. The suggestion is that when the splitting coefficient of the pressures $\Delta$ becomes $>1$, the differences in the pressures can not be neglected and a different approach must be used to study the structure of the star.

Our work consisted in taking a different symmetry as a starting point for the description of the structure of the star. The magnetic field fixes a preferred direction in space, for this reason we consider that an axisymmetric geometry is more ``natural'' in a magnetized system. By adopting a cylindrical symmetric metric we have solved Einstein equations and found the mass per unit length. We have obtained a regular behavior of the metric coefficients inside the star and a physically consistent dependence of the pressures with the radii.

Taking into account the pressure anisotropy due to a magnetic field leads to {\it higher} maximum masses in the theoretical stellar sequences, and to the existence of a critical field ($B_{c}\sim1.8\times10^{18}$ G) beyond which there are no equilibrium configurations. This critical field is essentially (up to a small factor) the one obtained based on the scalar virial theorem fulfillment $B_{\text{max}}\sim10^{18}$ G.

Our main simplification in this model is that we have taken that all the variables as being dependent just on the perpendicular (equatorial) radius, this allowed us to have a more tractable system of differential equations, although as a result we can not compute accurately the physical quantities unless the dependence with both $(r,\,z)$ of the variables of the problem is kept. Nevertheless,the model confirms the intuitive idea of the existence of a maximum magnetic field for which the star may undergo an anisotropic collapse due to a magnetic instability. We conjecture that a more accurate scheme should render slightly different values, but the same qualitative behavior found above.

\emph{Our model has been applied \cite{letter} to obtain the onset of the instability for magnetic white dwarfs, recently considered in several papers \cite{hindues}.}

\section*{Acknowledgments}
The authors  thanks to H.~Quevedo, R.~Pican\c{c}o, D.~A.~Foga\c{c}a, B.~Franzon and J. Rueda for fruitful comments and discussions. The work of A.P.M. and D.M.P. have been supported  under the grant CB0407 and the ICTP Office of External Activities through NET-35. D. M. P acknowledgment the fellowship CLAF-ICTP  and also thanks to IAG-USP for the hospitality. J.E.H. wishes to thank the financial support of the CNPq and FAPESP Agencies (Brazil). A.P.M thanks the hospitality and support of the International Center for Relativistic Astrophysics Network,  specially to Prof Remo Ruffini where this paper has been finished.


\end{document}